%% file: main.tex
\documentclass[sigconf, nonacm, screen]{acmart}

\AtBeginDocument{%
  }

\usepackage{enumitem}
\usepackage{graphicx}
\usepackage{subcaption}
\usepackage{float}
\usepackage{tikz}

\usepackage{soul} % for \st

\setlist[itemize]{align=parleft,left=0pt..1em}
\begin{document}

\title{Reliable Replication Protocols on SmartNICs}

\author{M.R. Siavash Katebzadeh}
\email{m.r.katebzadeh@ed.ac.uk}
\affiliation{%
  \institution{University of Edinburgh}
  \city{Edinburgh}
  \country{United Kingdom}
}
\author{Antonios Katsarakis}
\email{antonios.katsarakis@huawei.com}
\affiliation{%
  \institution{Huawei Research}
  \city{Edinburgh}
  \country{United Kingdom}
}

\author{Boris Grot}
\email{boris.grot@ed.ac.uk}
\affiliation{%
  \institution{University of Edinburgh}
  \city{Edinburgh}
  \country{United Kingdom}
}

\renewcommand{\shortauthors}{Katebzadeh et al.}
\newcommand*\circled[1]{\raisebox{.5pt}{\textcircled{\raisebox{0pt} {\footnotesize#1}}}}
\newcommand*\circledblack[1]{%
  \tikz[baseline=(char.base)]{
    \node[shape=circle, draw=black, fill=black, text=white, inner sep=1pt] (char) {\footnotesize #1};
  }%
}

\newif\ifshowcomment
\showcommenttrue
% \showcommentfalse

\ifshowcomment

\newcommand{\todo}[1]{\noindent\textsf{\color{red}{[{todo: \it #1}]}}}
\newcommand{\new}[1]{\textcolor{magenta}{#1}} % Added new texts
\newcommand{\modtext}[1]{\textcolor{blue}{#1}}  % Modified texts
\newcommand{\boris}[1]{\noindent\textsf{\color{magenta}{[Boris: {\it#1}]}}}
\newcommand{\siavash}[1]{\noindent\textsf{\color{magenta}{[Siavash: {\it#1}]}}}
\newcommand{\antonis}[1]{\noindent\textsf{\color{blue}{[Antonis: {\it#1}]}}}
\newcommand{\ak}[1]{\noindent\textsf{\color{orange}{[ak: {\it#1}]}}}
\else
\newcommand{\new}[1]{#1}
\newcommand{\modtext}[1]{#1}
\newcommand{\todo}[1]{}
\newcommand{\boris}[1]{}
\newcommand{\siavash}[1]{}
\newcommand{\antonis}[1]{}
\newcommand{\antonis}[1]{}
\newcommand{\ak}[1]{}
\fi

\begin{abstract}
  \input{0_abstract}
\end{abstract}

\maketitle

\input{1_introduction}

\input{2_background.tex}

\input{3_motivation.tex}
\input{4_relatedwork.tex}
\input{5_design.tex}
\input{6_implementation.tex}
\input{7_conclusion.tex}
\bibliographystyle{acm}
\bibliography{master}
\end{document}

%% file: 0_abstract.tex
Today's datacenter applications rely on datastores that are required to provide high availability, consistency, and performance.
To achieve high availability, these datastores replicate data across several nodes.
Such replication is managed through a reliable protocol designed to keep the replicas consistent using a consistency model, even in the presence of faults.
For several applications, strong consistency models are favored over weaker consistency models, as the former guarantee a more intuitive behavior for clients.
Furthermore, to meet the demands of high online traffic, datastores must offer high throughput and low latency.

However, delivering both strong consistency and high performance simultaneously can be challenging. Reliable replication protocols typically require multiple rounds of communication over the network stack, which introduces latency and increases the load on network resources.
Moreover, these protocols consume considerable CPU resources, which impacts the overall performance of applications, especially in high-throughput environments.

In this work, we aim to design a hardware-accelerated system for replication protocols to address these challenges.
We approach offloading the replication protocol onto SmartNICs, which are specialized network interface cards that can be programmed to implement custom logic directly on the NIC.
By doing so, we aim to enhance performance while preserving strong consistency, all while saving valuable CPU cycles that can be used for applications' logic.

%%% Local Variables:
%%% mode: LaTeX
%%% TeX-master: "./main"
%%% End:

%% file: 1_introduction.tex
\section{Introduction}

Modern applications and cloud services rely on distributed datastores to keep data safe and accessible.
Distributed datastores must balance the trade-off between consistency and performance.
  Often, they compromise on the strength of the consistency guarantees for performance.
Weak consistency models, such as eventual~\cite{petersen1996bayou,mahajan2011depot}, timeline~\cite{ahamad1995causal}, snapshot~\cite{attiya1995atomic} and causal~\cite{li1989memory} consistency, favor performance by
relaxing consistency;
  However, these weakly consistent models can be challenging to work with, as they often require developers to write custom code to achieve stronger semantics~\cite{vogels2009eventually}.
To overcome these limitations and improve programmability, coordination services such as ZooKeeper~\cite{huntZooKeeperWaitfreeCoordination} and Chubby~\cite{burrows2006chubby}, along with geo-replicated databases such as Spanner~\cite{corbett2013spanner} support strong consistency guarantees using algorithms such as variants of Paxos~\cite{lamport1998parttime, marandi2010ringpaxos, whittaker2020matchmaker, lamportVerticalPaxosPrimarybackup2009, lamport2001paxos, li2016just, whittaker2021scaling, RyabininGS24}.

However, while strongly consistent approaches are more desirable for correctness and programmability, they come at a performance cost.
  In many of strong consistency models, all {\ttfamily write} operations are serialized at a special node, generally referred to as the leader, which severely limits throughput in write-heavy workloads.
Some systems attempt to address this limitation.
For example, Attiya et al.~\cite{attiya1995sharing} propose a solution that achieves strong consistency even in the presence of failures, without relying on consensus algorithms like Paxos to determine the order of {\ttfamily writes}.
However, in their protocol, each {\ttfamily read} operation still requires communication with a quorum of replicas, which can significantly reduce {\ttfamily read} throughput~\cite{vasilisgavrielatosantonioskatsarakisOdysseyImpactModern2019}.

More recent works, including Microsoft's FaRM~\cite{dragojevicFaRMFastRemote2014} and several systems inspired by it~\cite{burkePRISMRethinkingRDMA2021}, have shown that it is possible to achieve both strong consistency and high performance.
These systems store data entirely in memory and leverage high-end networking technologies such as RDMA~\cite{Aguilera2019} to avoid the traditional bottlenecks associated with storage and TCP/IP networking stack in strongly consistent systems.
However, even in these systems, the CPU remains the limiting factor, as it is responsible for managing complex protocols and processing replication messages on each replica.

Our preliminary results show that across a variety of state-of-the-art replication protocols, between 15\% to 49\% of the CPU cycles are spent on network operations.
Moreover, up to 68\% of the CPU cycles are used by the replication operations.
With Moore's law slowing~\cite{leisersonTheresPlentyRoom2020}, the prospects for significant future improvements in CPU performance are limited;
therefore, to scale the performance of strongly consistent protocols, developing specialized hardware is becoming a reasonable option.

An increasingly popular approach in modern computing is the use of accelerators to offload workloads from CPU, thus, freeing up CPU cycles for other tasks.
In addition to high-speed data transfer using RDMA, network components like programmable switches and SmartNICs are emerging as key examples of such accelerators.
These accelerators offload parts of the network stack to enable the processing of data as it traverses the network;
therefore, they can potentially allow distributed systems to shift computation away from the CPU.
While recent work shows that security-related tasks (e.g., encryption/decryption), compression,  and tenant isolation in datacenters can be executed efficiently~\cite{10.1145/3627703.3650071, 10.1145/3387514.3405895, khalilov2024osmosis}, there is currently limited research demonstrating the offloading capabilities of these accelerators for replication protocols.

This paper proposes a new system called {\em Chaapar} that offers hardware-accelerated replication protocols using SmartNICs.
  At its core, Chaapar aims to reduce the CPU overhead of replication protocols on the host by offloading replication operations to the SmartNIC, thereby freeing up CPU cycles.
  By implementing key replication logic directly on the SmartNIC, Chaapar minimizes communication overhead and latency between replicas.
  Moreover, Chaapar integrates seamlessly with existing state-of-the-art datastores and cloud storage services while running them on the host.
  The system introduces an innovative architecture where SmartNICs cache data to minimize the PCIe overhead between the host and SmartNIC, thus optimizing the latency.
  With this architecture, Chaapar addresses the issues in prior work (Section~\ref{sec:relatedwork}): its scalability is not limited by the available memory on the SmartNIC, as it integrates with the host’s memory to handle larger datastores efficiently.
  We put particular effort into the design of Chaapar with these innovations to make it agnostic to different replication protocols, tailored towards multi-threaded, RDMA-enabled, in-memory, replicated datastores, and practical to use in datacenter deployments.

  The main contributions of this paper are the followings:
  \begin{itemize}
    \item We show that replication protocols, regardless of their design, incur significant CPU overheads that can limit their performance.
    \item We present {\em Chaapar}, a hardware-accelerated system for replication protocols that offloads replication operations onto SmartNICs.
      With its design, Chaapar optimizes communication both between replicas and between the host and the SmartNIC, all while freeing up CPU cycles on the host.
\end{itemize}
The rest of the paper is organized as follows: Section~\ref{sec:background} provides the necessary background.
Section~\ref{sec:motivation} evaluates the CPU usage of various replication protocols and identifies the sources of overhead.
Section~\ref{sec:relatedwork} reviews key prior work in the field, and highlights their limitations and challenges.
Section~\ref{sec:design} introduces the design of Chaapar, our hardware-accelerated replication protocol system.
Finally, Section~\ref{sec:impl} discusses the current status of our work, and outlines progress, challenges and future directions.

%%% Local Variables:
%%% mode: LaTeX
%%% TeX-master: "./main"
%%% End:

%% file: 2_background.tex
\section {Replicated Key-Value Stores} \label{sec:background}
Key-Value stores (KV-stores) are the backbone of data storage systems, e.g., databases.
A KV-store stores data as a collection of key-value pairs, generally in hashtables or LSM trees, and provides a {\ttfamily read}/{\ttfamily write} API that enables clients to perform operations on the stored data.
KV-stores are widely used as either primary stores, for example in Redis~\cite{redis} and RAMCloud~\cite{ousterhout15_ramcl_storag_system}, or caches in database systems such as Memcached~\cite{xu14_charac_faceb_memcac_workl} and disk-based datastores such as LevelDB and RocksDB~\cite{caoCharacterizingModelingBenchmarking}.

KV-stores are often replicated across multiple servers, usually ranging from 3 to 7 instances, known as the replication degree, to increase throughput and provide data availability in the presence of faults~\cite{katsarakisHermesFastFaultTolerant2020}.
The replication degree presents a trade-off between cost and fault tolerance: while adding more replicas enhances fault tolerance, it also raises the overall deployment costs and may negatively impact performance.
Clients connect to these replicated KV-stores through sessions.
The order of requests within each session is defined as the session order.
To ensure that concurrent accesses to KV-stores operate as expected, it is essential to maintain the replicas consistent.

The consistency model refers to the relationship between all replicas towards reflecting a coherent view of the data to all clients.
Weaker consistency models are known for their high performance by allowing more flexible data access patterns in distributed systems.
However, this flexibility often complicates programming, and developers need to manage synchronization explicitly.
In contrast, stronger consistency models strive to create the illusion of accessing data on a single server, meaning that all operations happen in a globally agreed order.
While these models provide predictable behavior and simplify programming, they may reduce the overall system performance due to the need for more coordination among replicas.
In this work, we focus on strong consistency models such as Sequential Consistency and Linearizability~\cite{attiya1994sequential}.

In a replicated KV-store, a consistency model is enforced through replication protocols~\cite{palmieri2011osare, katsarakisHermesFastFaultTolerant2020, huntZooKeeperWaitfreeCoordination, jhaDerechoFastState2019, enes2021efficient, lamport2001paxos}, which manage the coordination between replicas and perform data replication.
Replication protocols generally rely on consensus algorithms to achieve agreement among replicas on the order of operations.
These algorithms require multiple rounds of communication to reach an agreement.
Decisions in consensus algorithms are constrained by the latency of network round-trip times.

Achieving high performance while maintaining strong consistency and fault tolerance is a well-known challenge for replication protocols.
In the context of the KV-store, high performance is generally defined as low latency and high throughput.
The main two KV-store operations, {\ttfamily read} and {\ttfamily write}, require different optimizations and design considerations to provide high performance.

\noindent $\succ${\em Read:} To achieve high performance for {\ttfamily reads}, it is crucial to serve {\ttfamily read} operations from any replica, i.e., {\ttfamily reads} are load-balanced.
  Load-balancing {\ttfamily reads} remains challenging for many replication protocols.
  Protocols like Paxos generally require communication between replicas to agree on the correct read value.
  Other protocols, such as Primary-backup~\cite{lamportVerticalPaxosPrimarybackup2009}, enforce that only one replica can handle {\ttfamily reads} for a key.
  Similarly, Chain-Replication (CR)~\cite{renesseChainReplicationSupporting2004} protocol serves {\ttfamily read} operation at a node referred to as the tail node.

\noindent $\succ${\em Write:} Achieving high performance for {\ttfamily write} operations is even more difficult than for {\ttfamily reads}.
  A replication protocol requires the following properties to deliver high performance {\ttfamily writes}.
\begin{itemize}
\item Inter-key concurrency: Independent {\ttfamily writes} on different keys should proceed in parallel to enable multi-threaded execution.
  ZAB~\cite{junqueiraZabHighperformanceBroadcast2011}, for instance, serializes all {\ttfamily writes} through the leader, which limits concurrency.
\item Fast coordination: Performing {\ttfamily write} operations requires coordination among replicas to agree on the order of updates visible to the programmer.
  The coordination can generate many network activities among replicas.
Traditional networking stacks, such as TCP/IP, are not optimized for low latency or specific communication patterns, which can cause agreement protocols to become a bottleneck in a replicated KV-store.
Moreover, these stacks rely on CPU resources and consume CPU cycles that could otherwise be used to handle application tasks.
The increased CPU consumption can negatively impact the overall performance of running applications.
Recent work shows that the use of high-end networking technologies may enhance the performance of replication protocols -- especially on {\ttfamily writes}~\cite{vasilisgavrielatosantonioskatsarakisOdysseyImpactModern2019,katsarakisHermesFastFaultTolerant2020}.

\end{itemize}

%%% Local Variables:
%%% mode: LaTeX
%%% TeX-master: "./main"
%%% End:

%% file: 3_motivation.tex
\section {CPU Overheads of Replication Protocols} \label{sec:motivation}

Mainstream replication protocols typically follow a leader-based design, where a single node coordinates all {\ttfamily write} operations.
Protocols such as MultiPaxos~\cite{lamport2001paxos}, Raft~\cite{ongaroSearchUnderstandableConsensus2014}, ZAB, CHT~\cite{chandra2016algorithm}, and Primary-backup rely on a specific replica, generally referred to as a leader, to manage replication and ensure consistency.
While effective, such a centralized approach can create bottlenecks, particularly under high load.
To mitigate this limitation, some replication protocols distribute request handling across replicas.
For instance, in CR, which forms a chain of replicas, {\ttfamily write} requests are directed to the head node while {\ttfamily reads} are processed at the tail; potentially reducing contention on a single coordinator.
CRAQ~\cite{terraceObjectStorageCRAQ2009} improves over CR by allowing {\ttfamily reads} from all replicas, reducing {\ttfamily read} latency, but the head node must still initiate all {\ttfamily writes}.
Other protocols, such as Hermes~\cite{katsarakisHermesFastFaultTolerant2020}, AllConcur~\cite{pokeAllConcurLeaderlessConcurrent2017}, Tempo~\cite{enes2021efficient},  and Derecho~\cite{jhaDerechoFastState2019}, take decentralization further by allowing any replica to handle {\ttfamily write} requests and spread the workload more evenly.

In this section, we demonstrate that despite these design differences, all replication protocols impose significant CPU overhead.
Coordination, message processing, and consistency enforcement consume valuable CPU cycles.
We analyze how replication protocols, regardless of their architectural choices, contribute to CPU resource contention and explore a potential solution to alleviate this burden.

\begin{figure}[t]

  \centering
  \includegraphics[width=0.41\textwidth]{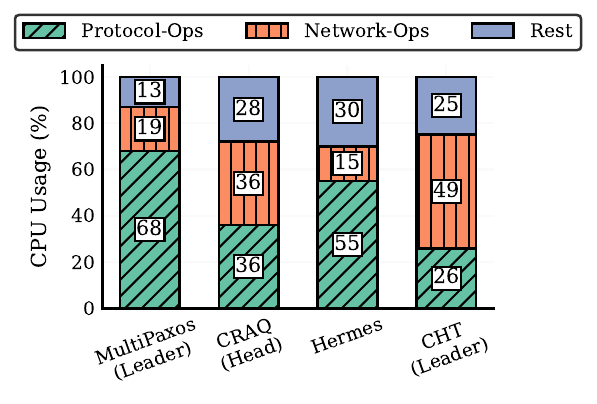}
\vspace{-1.8em}
   \caption{CPU usage breakdown across replication protocols.}
   \label{fig:cpu_usage}
\vspace{-1.5em}
\end{figure}

We profile four replication protocols: MultiPaxos, CRAQ, Hermes, and CHT.
To keep the comparison fair, we use the Odyssey framework~\cite{vasilisgavrielatosantonioskatsarakisOdysseyImpactModern2019}.
Odyssey implements these protocols in a multi-threaded design, and leverages high-end RDMA-enabled NICs to optimize network operations.
In our experiment, we run each protocol inside a cluster of five replicas.
These replicas are connected through a 56Gbps InfiniBand switch.
We measure the number of CPU cycles spent on three categories of operations:
\circled{1} Protocol-specific operations, such as invalidation, write commit, write ordering, versioning (timestamping), etc.;
\circled{2} Network-specific operations, such as broadcasting and multicasting messages as well as receiving and parsing them, and
\circled{3} the Rest category, which includes client operations such as sending {\ttfamily read} and {\ttfamily write} queries.
Our experiment generates a uniform {\ttfamily read}/{\ttfamily write} trace with 20\% to 80\% ratio.
The KV-store is pre-populated with one million KV pairs, all replicated on all nodes.
Similar to prior studies, the size of keys and values are 8 and 32 bytes, respectively~\cite{alimadadiWaverunnerElegantApproach}.
We set the failure rate to zero to observe the CPU usage in the absence of failure.

Figure~\ref{fig:cpu_usage} illustrates the CPU breakdown across different protocols.
Note that the figure demonstrates CPU breakdown for leaders of MultiPaxos and CHT, as well as the head replica in CRAQ protocols.
While we show the profiling results of a random node when running the Hermes protocol, which is decentralized, our results show that the CPU usage remains consistent across different replicas when the trace follows a uniform distribution.

We observe that a majority of the CPU cycles are used by protocol- and network-specific operations of these protocols.
As the figure shows, CPU usage for protocol-specific operations varies from 26\% to 68\%.
The figure also shows that these protocols spend between 15\% and 49\% of the CPU cycles on network-specific operations.

{\em Where does the overhead come from?} We identify three main sources of overhead for replication protocols:

    \noindent {\bf \circledblack{1} Communication:} Broadcasting/multicasting of messages such as proposal, invalidation, validation, etc., to all replicas incurs unavoidable overhead.

    \noindent {\bf \circledblack{2} Expensive PCIe transactions:}
      To reduce network latency, the implementation of these protocols in the Odyssey framework uses RDMA and relies on spin loops to fetch messages as quickly as possible.
      While RDMA provides low-latency networking, transmitting the messages from NIC to host (and vice versa) over PCIe is expensive and increases end-to-end latency.
      Using spin loops may reduce latency, but it does not eliminate the inherent PCIe overhead and increases CPU usage.
     One-sided RDMA operations are often seen as an attractive solution for bypassing remote CPU involvement;
     however, prior works such as Odyssey have demonstrated that overall, they increase the CPU load on the initiator and limit batching over the network and PCIe, leading to higher overhead.

    \noindent {\bf \circledblack{3} Concurrency ordering:} Different replication protocols use approaches to enforce either total order  (e.g., ZAB, MultiPaxos, Raft, Derecho, AllConcur, and Mencius) or per-key order (e.g., CHT, CRAQ, Classic Paxos and Hermes).
      Managing {\ttfamily write} ordering adds software overhead to the system.

\noindent\textbf{Insight:}
Hardware accelerators are valid candidates to implement complex replication protocols and reduce network latency by offloading the operations.
Implementing the protocols on hardware accelerators can save CPU cycles spent on broadcasting/multicasting messages as well as concurrency ordering.

%%% Local Variables:
%%% mode: LaTeX
%%% TeX-master: "./main"
%%% End:

%% file: 4_relatedwork.tex
% !TEX root = ../main.tex

\vspace{-1em}
\section{Replication Protocols on SmartNICs}\label{sec:relatedwork}

In addition to high-end RDMA-enabled networking stacks, modern networks increasingly feature programmable components like switches and NICs that provide on-device computation.
These components enable the manipulation of data as it transits the network and allow distributed systems to offload computations and improve performance.
SmartNICs, like Nvidia's BlueField~\cite{nvidia_bluefield2}, Huawei's IN5500~\cite{huawei_in550}, Broadcom's Stingray~\cite{broadcom_stingray}, Marvell's LiquidIO~\cite{marvell2021liquidio}, and Netronome's Agilio~\cite{netronome2021agilio}, include onboard memory ranging from 8 to 32 GB, along with different computation units, such as SoC or FPGA, integrated into NICs.
These NICs allow developers to access the computation units to make it possible to offload customized computation logic.

The advent of SmartNICs has sparked some interest in their utilization within the distributed systems community.
SmartNICs have the potential to enhance the performance and efficiency of KV-store operations, particularly in the context of consensus and replication protocols.
Recent works, such as ZABFPGA~\cite{istvanConsensusBoxInexpensive2016} and Waverunner~\cite{alimadadiWaverunnerElegantApproach},  are generally centered around FPGA-based SmartNICs to implement custom logic for protocol- and network-specific operations to improve throughput and reduce CPU bottlenecks.

\noindent{\bf $\blacktriangleright$ ZABFPGA: } offloads the ZAB protocol onto an FPGA-based SmartNIC, integrating a network stack, atomic broadcast module, and KV-store directly on the FPGA.
The network stack is based on TCP and optimized for low latency by employing dataflow pipelines, and tailored for datacenters.
The atomic broadcast module replicates {\ttfamily write} requests so that all nodes receive the same sequence of operations, and handles {\ttfamily reads} locally.
By integrating the KV-store with the atomic broadcast module, ZABFPGA eliminates PCIe overhead.

{\bf Limitations:}
  While integrating the KV-store with the atomic broadcast unit enhances performance and concurrent data access, ZABFPGA’s hardware-only approach presents significant challenges:

 \noindent $\succ$  {\em Scalability limitation:} Due to the limited DRAM capacity of SmartNICs (usually up to 32 GB), ZABFPGA cannot scale effectively and is unable to handle KV-stores with hundreds of gigabytes, which severely limits its applicability for large-scale systems.

 \noindent $\succ$  {\em Challenges of FPGA development:} FPGA implementation of complex replication protocols requires specialized expertise, including the development of a hardware network stack and handling operations like leader election and failure recovery.
  These tasks are error-prone, even in software, and addressing all corner cases in hardware is notoriously difficult, making the design less suitable for practical deployment.

\noindent{\bf $\blacktriangleright$ Waverunner:} attempts to address the scalability issue of ZABFPGA by using a hybrid approach that combines hardware and software components.
  Unlike ZABFPGA's hardware-only design, Waverunner offloads the network-related parts of the Raft protocol to the FPGA-based SmartNIC, while the rest of the protocol runs in software on the host.
  The hardware component handles common Raft messages, bypassing the kernel network stack, while more complex operations such as leader election and failure recovery remain in the software.
  This hybrid approach makes the system more practical by allowing fallback to a software-only mode when necessary.

{\bf Limitations:} \\
\noindent $\succ$ {\em PCIe overhead:} Unlike ZABFPGA, Waverunner runs a significant portion of Raft protocol as well as the KV-store on the host, requiring expensive communication between the host and the SmartNIC over PCIe, which introduces additional latency.

\noindent $\succ$ {\em Challenges of FPGA development:}
  While Waverunner offers a clean design by offloading only some parts of the Raft protocol, it still faces a similar practical issue as ZABFPGA: implementing even small parts of the protocol on FPGA is challenging.

\noindent{\bf Takeway: }
Both ZABFPGA and Waverunner offer innovative approaches by offloading replication protocols to SmartNICs, but they fail to fully address practical concerns for real-world deployments in datacenters.
ZABFPGA’s hardware-only approach struggles with FPGA development complexity and error-prone operations like leader election.
Furthermore, its reliance on the limited memory of SmartNICs hinders scalability, making it unsuitable for large-scale systems that require handling vast amounts of data; thus, difficult to deploy effectively in production environments.
Waverunner, despite its hybrid design, suffers from latency due to PCIe communication between the host and offloaded components.
Neither system fully leverages RDMA; instead, they use FPGA resources to implement custom network stacks, which is a waste of resources.

%%% Local Variables:
%%% mode: LaTeX
%%% TeX-master: "./main"
%%% End:

%% file: 5_design.tex
\section{Design Overview}\label{sec:design}

\subsection{Insights}
In this work, we take a pragmatic approach to design a system that can be effectively deployed in datacenters.
  Building on insights gained from studying related work in hardware-accelerated replication protocols, we find the followings:

\noindent{\bf Breaking the trade-off between consistency and performance:}
While achieving low latency and high throughput in strongly consistent systems is inherently challenging, our goal is to overcome this trade-off.
To achieve low latency, it is critical to minimize communication over PCIe, as it introduces a significant delay—at least 500ns round-trip latency—which impacts performance in latency-sensitive systems~\cite{katebzadehEvaluationInfiniBandSwitch2020}.
In systems like Waverunner, where much of the Raft protocol's logic is executed on the host, frequent communication between the host and SmartNIC over PCIe adds additional latency.
This constant data exchange over PCIe prevents the system from optimizing latency.
By offloading more processing to the SmartNIC, PCIe traffic can be reduced, leading to lower latency and improved performance.
Furthermore, since both Waverunner and ZABFPGA implement custom network stacks, they overlook the advanced RDMA features in SmartNICs.
RDMA can bypass the CPU, reduce memory copies, lower latency, and ease CPU load.
Additionally, RDMA’s batching capabilities can help reduce PCIe transactions, further optimizing PCIe overhead.
For high throughput, our system must be designed to easily adopt high-performance, leaderless replication protocols such as Hermes and AllConcur.
Leaderless protocols allow for better scalability by removing bottlenecks typically associated with leader-based designs and allowing concurrent {\ttfamily writes}.
Such a careful design allows the system to balance strong consistency with performance, breaking the traditional trade-off between the two.

\noindent{\bf Practical deployment for datacenter:}
While FPGAs offer flexibility and performance, they are challenging to deploy in real-world datacenter environments due to the specialized skills and experience required for their development.
Moreover, FPGA-based systems face longer development cycles and higher risks of design errors, making them unattractive for large-scale, production environments.
Additionally, a practical solution should also integrate easily with existing production-grade datastores, without requiring major changes or causing disruptions to ongoing operations.

\begin{figure}

  \centering
  \includegraphics[width=0.5\textwidth]{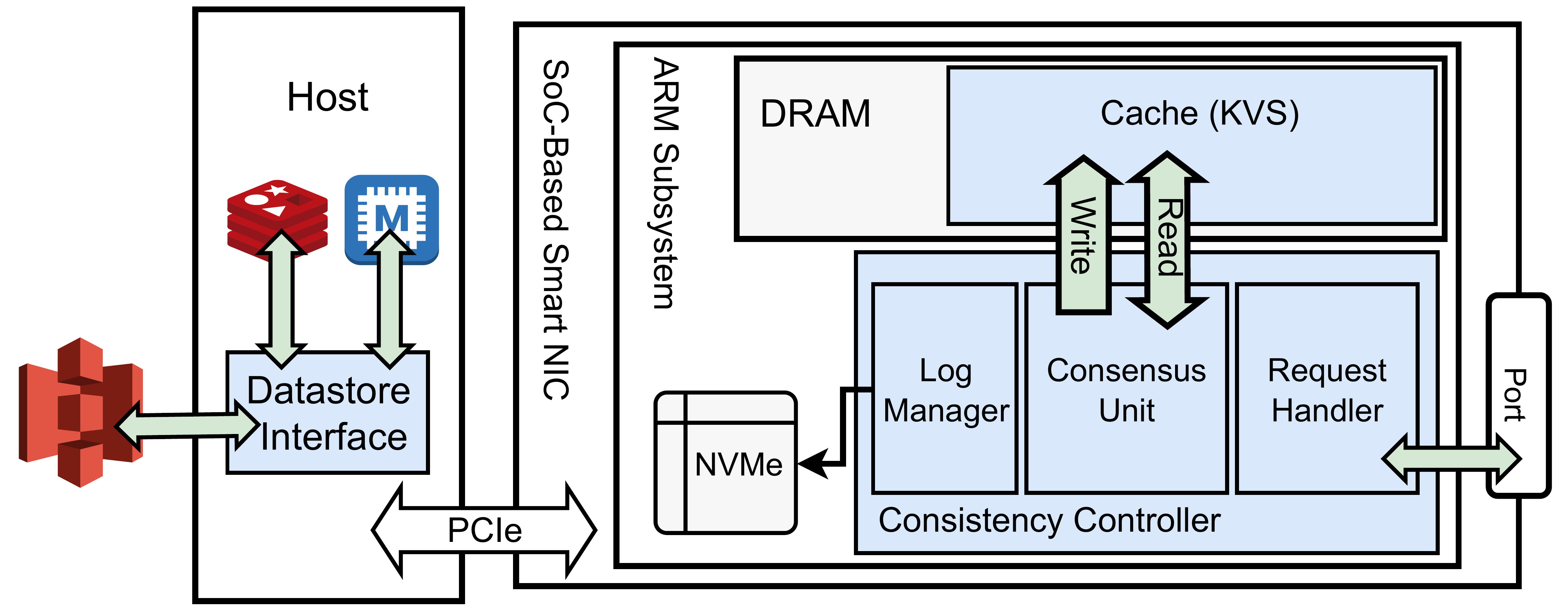}
% \vspace{-1.5em}
   \caption{Overview design of Chaapar}
   \label{fig:design}
   \vspace{-1em}
\end{figure}

\begin{figure*}[t]
    \centering
    \begin{subfigure}[t]{0.25\textwidth}
        \centering
        \includegraphics[width=\textwidth]{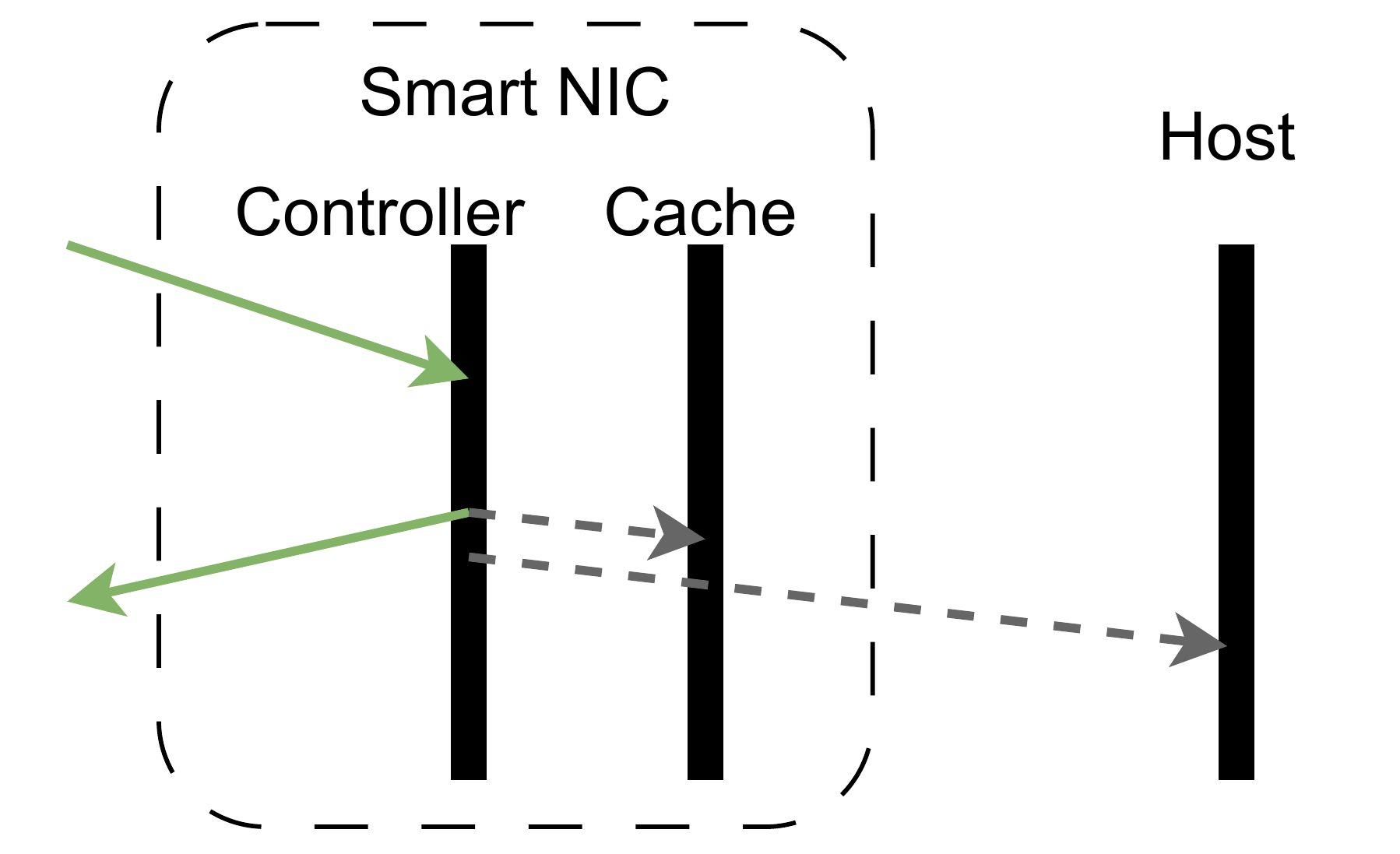}
        \caption{Write on fast path}
        \label{fig:write_fast}
    \end{subfigure}%
    \hfill
    \begin{subfigure}[t]{0.25\textwidth}
        \centering
        \includegraphics[width=\textwidth]{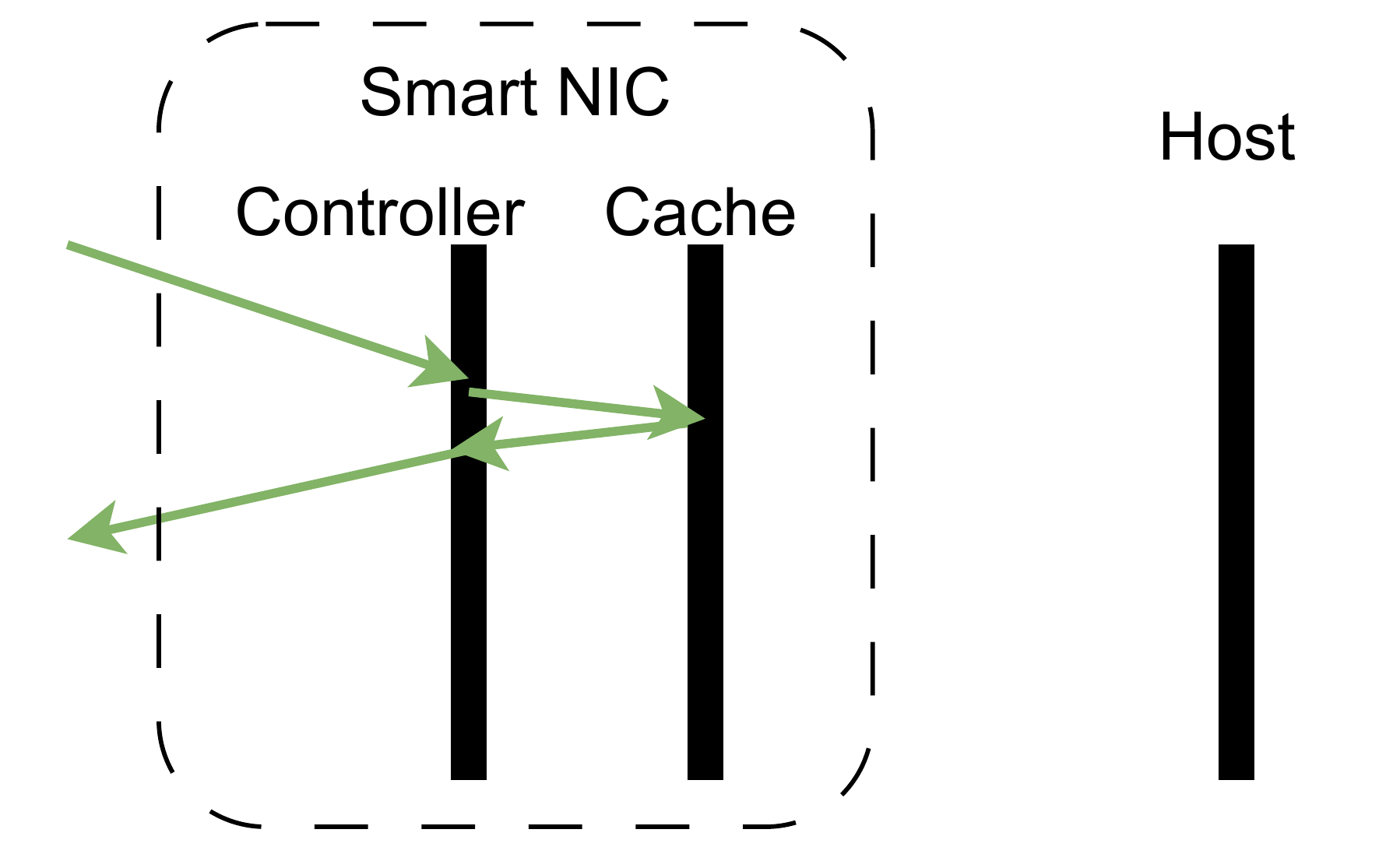}
        \caption{Read on fast path}
        \label{fig:read_fast}
    \end{subfigure}%
    \hfill
    \begin{subfigure}[t]{0.25\textwidth}
        \centering
        \includegraphics[width=\textwidth]{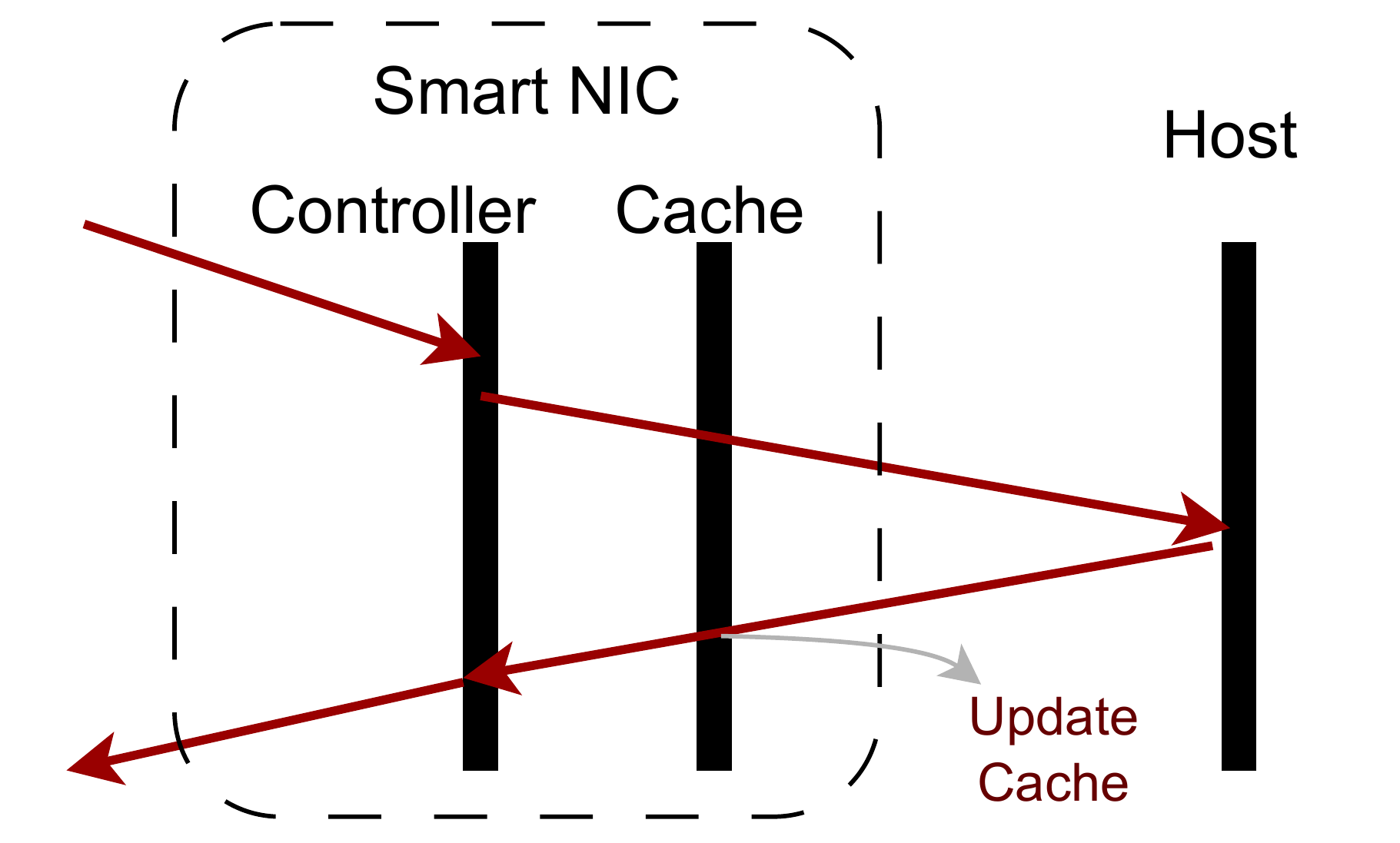}
        \caption{Read on slow path}
        \label{fig:read_slow}
    \end{subfigure}
\vspace{-1em}
    \caption{Overview of fast and slow paths in dual-path caching.}
    \label{fig:fast_slow_paths}

\vspace{-1em}
\end{figure*}

\subsection{Chaapar System}
The central hypothesis explored in this paper is that {\em strongly consistent replication protocols resemble cache coherence protocols} by sharing structural and operational similarities with them.
Cache coherence protocols, proven to be efficiently managed by hardware-based state machines, serve as a powerful design inspiration.
We aim to adapt these well-understood hardware-based coherence protocols to distributed systems at the datacenter scale.
To do so, we design {\em Chaapar}, a system that offers hardware-accelerated replication protocols using SmartNICs.
Chaapar is designed to offload the replication operations from the host CPU, free up precious CPU cycles, and improve latency.
Chaapar consists of three key innovations:
\circled{1}
Inspired by the design of modern CPUs, Chaapar implements a {\em consistency controller} on the SmartNIC of each replica to reduce communication overhead over the network and PCIe.
 In this design, the consistency controller manages the consistency between the datastore on the hosts or the cloud storage services.
\circled{2}
Chaapar maintains a {\em software-based data cache} on the SmartNICs to improve end-to-end latency.
To perform {\ttfamily read} and {\ttfamily write} operations, our system follows a {\em dual-path} approach:
 The fast path is used when the requested key exists in the software cache on the SmartNIC.
 On the slow path, the consistency controller fetches the required data from the datastore on the host using RDMA network operations to optimize network overhead.
\circled{3}
 Chaapar implements a {\em datastore interface} to seamlessly integrate with the state-of-the-art datastores and cloud storage services, and run them on the host CPUs of replicas.
Figure~\ref{fig:design} demonstrates Chaapar and its components cooperatively working to provide reliable replication protocol using SmartNICs.

\subsection{Consistency Controller}
At the core of Chaapar, there is a {\em consistency controller} – a hardware-accelerated protocol engine residing on the NIC – to perform replication- and network-specific operations of protocols to free the host CPU.
The consistency controller on the SmartNIC is responsible for enforcing consistency and comprises three main components:

\noindent{\bf Request Handler.} Request handler manages communication with other replicas for propagating {\ttfamily writes} and collecting acknowledgments, and also handles client requests asynchronously.
For the communication with other replicas, this unit leverages the RDMA engine on the SmartNIC to implement efficient multicast functionality.
It also provides several network interfaces, including Linux socket, gRPC, and RDMA to the clients and processes incoming client requests, and forwards them to the next component in the pipeline, the consensus unit.

\noindent{\bf Consensus Unit.} Responsible for implementing the consensus protocol, the consensus unit processes {\ttfamily read} and
{\ttfamily write} operations forwarded from the request handler.
The consensus unit deploys a digraph-based network overlay over the replicas to define the multicast pattern.
It leverages the overlay network to support leader-based, chained, and leaderless protocols.
In the current design, a leaderless approach is implemented that uses a 2-phase protocol to propagate updates to each of the replicas.
Thus, a {\ttfamily write} operation multicasts the new object in the 1st phase to all replicas and waits for acknowledgments from them; then, in the 2nd phase, it sends a commit message to all replicas.
The consensus unit provides fast local {\ttfamily reads}, i.e., it services the {\ttfamily read} requests without communicating with other replicas.
To ensure strong consistency, the consensus unit keeps track of uncommitted {\ttfamily writes} to block {\ttfamily read} requests to the same key with pending {\ttfamily writes}.
Transitions between states are triggered either by incoming requests from the request handler or by event timers, which are used to implement timeouts for tasks like detecting failed nodes.

\noindent{\bf Log Manager.} The log manager maintains an update log essential for fault tolerance and replaying operations in case of failures.
When the consensus unit receives a {\ttfamily write} request, the operation is appended to this log in an append-only manner.
Later, it reads out these operations and sends them as proposal commands to other replicas.
Upon committing an operation, an event is also added to the log.
In the event of failure, synchronization begins from the section of the log that holds the command entries to ensure consistency across replicas.
Additionally, the log can be compacted up to the last successful {\ttfamily write} to the datastore on the host; the datastore interface (Section~\ref{sec:interface}) signals the controller for each successful {\ttfamily write}, which can be leveraged by the controller for efficient log management and space optimization.

\subsection{Dual-Path Design Caching}
In our architecture, the datastore operates on the host.
In order to hide the latency of accessing the host through PCIe, the consistency controller maintains a DRAM-resident KV-store on the SmartNIC as a software-based data cache.
We design a {\em dual-path} mechanism to handle {\ttfamily read} and {\ttfamily write} requests to enable the system to dynamically switch between fast and slow paths based on data availability in the software cache.

\noindent{\bf Fast path.}
On the fast path, the access to the host is off the critical path; therefore PCIe latency does not contribute to the overall latency of the operation.

\underline{All {\ttfamily write} requests} are processed on the fast path.
As illustrated in Figure~\ref{fig:write_fast}, when the request handler receives a {\ttfamily write} request, it forwards it to the consensus unit, which commits the {\ttfamily write} (after communicating with other replicas), notifies the client, and caches the key-value pair on the SmartNIC.
In this design, updating the datastore on the host is batched and performed asynchronously, keeping host access off the critical path.

Similarly, when a {\ttfamily read} request results in a {\em cache hit}, it is also processed on the fast path, as shown in Figure~\ref{fig:read_fast}.
Here, the request handler forwards the request to the consensus unit, which checks for any uncommitted {\ttfamily writes} to the key.
If no uncommitted {\ttfamily write} exists, the request is serviced directly from the software cache.

\noindent{\bf Slow path.} If a {\ttfamily read} request results in a {\em cache miss}, the request is handled on the slow path.
In this scenario, the consistency controller communicates with the datastore on the host to retrieve the required data:
As depicted in Figure~\ref{fig:read_slow}, the request handler passes the {\ttfamily read} request to the consensus unit, which then fetches the data over PCIe from the host.
While servicing the client, the controller caches the key-value pair for future fast-path access.

This design leverages hardware acceleration on the NIC to offload consistency responsibilities from the host and minimizes the communication over PCIe to effectively balance performance with fault tolerance and enable high-speed, strongly consistent operations at scale.

\subsection {Datastore Interface} \label{sec:interface}
Unlike prior work, where the replication protocol is integrated with a simple KV-store, Chaapar is designed to seamlessly integrate with state-of-the-art and industry-standard KV-stores (e.g., Redis and Memcached) as well as cloud storage services (e.g., Amazon S3~\cite{amazon_web_services}).
This is achieved through the datastore interface.
The datastore interface is a key component running on the host that provides an abstraction layer for accessing the underlying datastore.
Its primary role is to manage and route requests from the consistency controller on the SmartNIC through RDMA.

This component serves the {\ttfamily read} and {\ttfamily write} operations, depending on the path.
On the fast path, the datastore interface handles requests when a {\ttfamily write} operation from the consistency controller must be applied directly to the datastore.
It ensures that the operation is durably committed to the storage backend and notifies the consistency controller upon a successful {\ttfamily write}.
Signaling the consistency controller enables the log manager to perform log compaction and release log space for future operations.

On the slow path, the datastore interface addresses cache misses encountered during {\ttfamily read} operations.
When the consistency controller identifies a cache miss for a requested key, it forwards the {\ttfamily read} request to the datastore interface, which retrieves the data from the datastore and returns it to the consistency controller.
This setup minimizes latency by keeping the {\ttfamily read} processing streamlined, avoiding direct host involvement unless necessary.

Additionally, the datastore interface functions as a flexible adapter for the datastore, providing an abstraction that allows the system to operate with various local
 or online datastore implementations without requiring modifications to the consistency controller.

Since {\ttfamily writes} to the datastore are off the critical path, the consistency controller can optimize performance by maintaining a {\ttfamily write} buffer on the SmartNIC.
When this buffer fills, a batch of {\ttfamily write} operations is sent to the datastore interface, reducing the frequency of individual messages.
With this design, PCIe communication overhead is well-managed;
for instance, the 32-byte header overhead for PCIe packets can be significant for small messages.
The batching strategy minimizes the effect of this overhead to further optimize the over the PCIe communication.

\subsection{Leveraging SoC-Based SmartNICs}
Chaapar explores SoC-based SmartNICs as an alternative to FPGA-based SmartNICs.
SoC-based SmartNICs offer several advantages that are aligned with our insights and make it more suitable for implementing replication protocols efficiently:

\noindent{\bf ARM cores: } SoC-based SmartNICs are generally equipped with ARM cores, thus we can offload replication protocols without requiring hardware description languages like those needed for FPGAs.
From the NIC's perspective, the ARM subsystem functions as a second full-fledged host with its own network interface.
This makes development more accessible and easier to debug and maintain.

\noindent{\bf Ready-to-use RDMA-enabled network stack:} Unlike some other solutions that require rebuilding a custom network stack, RDMA capabilities of SoC-based SmartNICs provide ultra-low latency and high bandwidth out of the box.
This allows us to take full advantage of its high-performance networking without reinventing the wheel, streamlining the implementation of the replication protocols.

\noindent{\bf Minimizing PCIe Transactions:} Programming on SoC-based SmartNICs is more straightforward than FPGAs and it enables us to implement most of the replication protocol logic directly on the SmartNIC.
This feature allows us to minimize PCIe communication, which typically introduces latency, and instead handle more processing on the SmartNIC itself; hence, we can free up host-side CPU cycles.

%%% Local Variables:
%%% mode: LaTeX
%%% TeX-master: "./main"
%%% End:

%% file: 6_implementation.tex
\section{Implementation \& Current Status} \label{sec:impl}

In this work, we have implemented the request handler and the datastore interface on the BlueField-3 card~\cite{burstein2021nvidia}, which is a SoC-based SmartNIC.
Our work is in progress and further development of the consensus unit and various optimizations remain to be implemented.
While offloading replication protocols to SmartNICs offers promising benefits in reducing host CPU overhead and improving network efficiency, several challenges must be addressed to fully realize the potential of this approach:

\noindent{\bf Limited compute power of SmartNIC SoCs:}
SmartNICs, like BlueField-3, are equipped with computation cores designed primarily for power efficiency rather than high performance.
As a result, their computational capabilities are significantly lower compared to modern CPUs used in hosts.
To overcome this limitation, it is crucial to maximize the use of specialized hardware components available on SmartNICs, such as DMA engines.
These components are well-suited to handle data transfers, allowing the ARM cores to focus on protocol-specific operations without becoming overloaded.

\noindent{\bf Challenges with implementing software cache on SmartNICs:}
Many state-of-the-art hashtables are heavily optimized for high-performance x86 CPUs, which benefit from faster memory access compared to the ARM cores found in BlueField-3.
This makes these designs unsuitable for the low-power ARM cores on SmartNICs, which lack the equivalent high-speed memory access.
Consequently, careful design of the request handler in the consistency controller is necessary to minimize contention and concurrent access to the shared software cache.
There is exciting potential for future research to develop efficient, concurrent hashtables specifically tailored for ARM cores, which could further enhance the performance of SmartNIC-based replication protocols.

However, these challenges are not inherent to the design or use of SmartNICs with ARM cores.
Each generation of SmartNICs is becoming increasingly powerful; for instance, the gap between the 2nd and 3rd generations of BlueField cards is already significant~\cite{michalowicz2023battle}.
Moreover, ARM cores themselves are not inherently unsuitable for such tasks.
For instance, Graviton ARM CPUs used in AWS have demonstrated substantial performance and power efficiency~\cite{awsgraviton4}.
Our design explores the use of SoC-based SmartNICs, and as the hardware continues to evolve, we expect these solutions to become more powerful, further closing the gap between SmartNICs and host CPUs, and making our system even more effective.

\noindent{\bf Supporting diverse replication protocols:}
To support multiple replication protocols and varying consistency guarantees, a flexible and modular architecture is required.
Different protocols, such as Raft or Hermes, have unique requirements for message sequencing, leader election (if needed), and failure recovery.
Implementing such diversity on SmartNICs, which are resource-constrained compared to host CPUs, presents challenges in designing general-purpose components that remain efficient across different use cases.

%%% Local Variables:
%%% mode: LaTeX
%%% TeX-master: "./main"
%%% End:

%% file: 7_conclusion.tex
\section{Conclusion}
This paper introduces Chaapar, a novel system that leverages SmartNICs to implement hardware-accelerated replication protocols.
Chaapar is designed to offload replication operations from the host CPU, freeing up valuable CPU cycles and enhancing overall system performance.
Inspired by modern CPU designs, Chaapar maintains a software-based data cache and a consistency controller on each replica's SmartNIC to minimize communication overhead across the network and PCIe.
Chaapar introduces three key innovations: offloading replication operations to SmartNIC's cores, seamless integration with state-of-the-art datastores and cloud storage services, and a dual-path approach to optimize read and write operations using RDMA.
This protocol-agnostic design demonstrates the potential of SmartNICs to enhance the efficiency and flexibility of reliable replication protocols.

%%% Local Variables:
%%% mode: LaTeX
%%% TeX-master: "./main"
%%% End:

%% file: main.bbl
\begin{thebibliography}{10}

\bibitem{amazon_web_services}
Amazon web services, cloud object storage s3.
\newblock \url{http://s3.amazonaws.com}.

\bibitem{awsgraviton4}
Amazon web services, graviton4 arm-based processors, 2024.
\newblock Available at: \url{https://aws.amazon.com/ec2/graviton/}.

\bibitem{Aguilera2019}
{\sc Aguilera, M.~K., {Ben-David}, N., Guerraoui, R., Marathe, V., and
  Zablotchi, I.}
\newblock The impact of {{RDMA}} on agreement.
\newblock {\em Proceedings of the Annual ACM Symposium on Principles of
  Distributed Computing\/} (2019), 409--418.

\bibitem{ahamad1995causal}
{\sc Ahamad, M., Neiger, G., Burns, J.~E., Kohli, P., and Hutto, P.~W.}
\newblock Causal memory: Definitions, implementation, and programming.
\newblock {\em Distributed Computing 9}, 1 (1995), 37--49.

\bibitem{alimadadiWaverunnerElegantApproach}
{\sc Alimadadi, M., Mai, H., Cho, S., Ferdman, M., Milder, P., and Mu, S.}
\newblock Waverunner: {{An Elegant Approach}} to {{Hardware Acceleration}} of
  {{State Machine Replication}}.

\bibitem{attiya1995sharing}
{\sc Attiya, H., Bar-Noy, A., and Dolev, D.}
\newblock Sharing memory robustly in message-passing systems.
\newblock {\em Journal of the ACM 42}, 1 (jan 1995), 124--142.

\bibitem{attiya1995atomic}
{\sc Attiya, H., Herlihy, M., and Rachman, O.}
\newblock Atomic snapshots using lattice agreement.
\newblock {\em Distributed Computing 8\/} (1995), 121--132.

\bibitem{attiya1994sequential}
{\sc Attiya, H., and Welch, J.~L.}
\newblock Sequential consistency versus linearizability.
\newblock {\em ACM Transactions on Computer Systems (TOCS) 12}, 2 (1994),
  91--122.

\bibitem{broadcom_stingray}
{\sc {Broadcom}}.
\newblock {Stingray SmartNIC Adapters and IC}.
\newblock
  \url{https://www.broadcom.com/products/ethernet-connectivity/smartnic}.

\bibitem{burkePRISMRethinkingRDMA2021}
{\sc Burke, M., Dharanipragada, S., Joyner, S., Szekeres, A., Nelson, J.,
  Zhang, I., and Ports, D. R.~K.}
\newblock {{PRISM}}: {{Rethinking}} the {{RDMA Interface}} for {{Distributed
  Systems}}.
\newblock In {\em Proceedings of the {{ACM SIGOPS}} 28th {{Symposium}} on
  {{Operating Systems Principles}}\/} (Virtual Event Germany, Oct. 2021), ACM,
  pp.~228--242.

\bibitem{burrows2006chubby}
{\sc Burrows, M.}
\newblock The chubby lock service for loosely-coupled distributed systems.
\newblock In {\em Proceedings of the 7th symposium on Operating systems design
  and implementation\/} (2006), pp.~335--350.

\bibitem{burstein2021nvidia}
{\sc Burstein, I.}
\newblock Nvidia data center processing unit (dpu) architecture.
\newblock In {\em 2021 IEEE Hot Chips 33 Symposium (HCS)\/} (2021), pp.~1--20.

\bibitem{caoCharacterizingModelingBenchmarking}
{\sc Cao, Z., Dong, S., Vemuri, S., and Du, D. H.~C.}
\newblock Characterizing, {{Modeling}}, and {{Benchmarking RocksDB Key-Value
  Workloads}} at {{Facebook}}.

\bibitem{chandra2016algorithm}
{\sc Chandra, T.~D., Hadzilacos, V., and Toueg, S.}
\newblock An algorithm for replicated objects with efficient reads.
\newblock In {\em Proceedings of the 2016 ACM Symposium on Principles of
  Distributed Computing\/} (2016), pp.~325--334.

\bibitem{corbett2013spanner}
{\sc Corbett, J.~C., Dean, J., Epstein, M., Fikes, A., Frost, C., Furman,
  J.~J., Ghemawat, S., Gubarev, A., Heiser, C., Hochschild, P., et~al.}
\newblock Spanner: Google’s globally distributed database.
\newblock {\em ACM Transactions on Computer Systems (TOCS) 31}, 3 (2013),
  1--22.

\bibitem{dragojevicFaRMFastRemote2014}
{\sc Dragojevi{\'c}, A., Narayanan, D., Castro, M., and Hodson, O.}
\newblock {{FaRM}}: {{Fast Remote Memory}}.
\newblock {\em Nsdi'14\/} (2014), 401--414.

\bibitem{enes2021efficient}
{\sc Enes, V., Baquero, C., Gotsman, A., and Sutra, P.}
\newblock Efficient replication via timestamp stability.
\newblock In {\em Proceedings of the Sixteenth European Conference on Computer
  Systems\/} (2021), pp.~178--193.

\bibitem{10.1145/3387514.3405895}
{\sc Grant, S., Yelam, A., Bland, M., and Snoeren, A.~C.}
\newblock Smartnic performance isolation with fairnic: Programmable networking
  for the cloud.
\newblock In {\em Proceedings of the Annual Conference of the ACM Special
  Interest Group on Data Communication on the Applications, Technologies,
  Architectures, and Protocols for Computer Communication\/} (New York, NY,
  USA, 2020), SIGCOMM '20, Association for Computing Machinery, p.~681–693.

\bibitem{huawei_in550}
{\sc {Huawei}}.
\newblock {Huawei IN550 SmartNIC}.
\newblock \url{https://e.huawei.com/us/news/it/201810171443}, 2018.

\bibitem{huntZooKeeperWaitfreeCoordination}
{\sc Hunt, P., Konar, M., Junqueira, F.~P., and Reed, B.}
\newblock {{ZooKeeper}}: {{Wait-free}} coordination for {{Internet-scale}}
  systems.

\bibitem{istvanConsensusBoxInexpensive2016}
{\sc Istvan, Z., Sidler, D., Alonso, G., and Vukolic, M.}
\newblock Consensus in a {{Box}}: {{Inexpensive Coordination}} in {{Hardware}}.
\newblock In {\em 13th {{USENIX Symposium}} on {{Networked Systems Design}} and
  {{Implementation}} ({{NSDI}} 16)\/} (2016), pp.~425--438.

\bibitem{jhaDerechoFastState2019}
{\sc Jha, S., Behrens, J., Gkountouvas, T., Milano, M., Song, W., Tremel, E.,
  Renesse, R.~V., Zink, S., and Birman, K.~P.}
\newblock Derecho: {{Fast State Machine Replication}} for {{Cloud Services}}.
\newblock {\em ACM Transactions on Computer Systems 36}, 2 (Apr. 2019),
  4:1--4:49.

\bibitem{junqueiraZabHighperformanceBroadcast2011}
{\sc Junqueira, F.~P., Reed, B.~C., and Serafini, M.}
\newblock Zab: {{High-performance}} broadcast for primary-backup systems.
\newblock In {\em 2011 {{IEEE}}/{{IFIP}} 41st {{International Conference}} on
  {{Dependable Systems}} \& {{Networks}} ({{DSN}})\/} (Hong Kong, June 2011),
  IEEE, pp.~245--256.

\bibitem{katebzadehEvaluationInfiniBandSwitch2020}
{\sc Katebzadeh, M. R.~S., Costa, P., and Grot, B.}
\newblock Evaluation of an {{InfiniBand Switch}}: {{Choose Latency}} or
  {{Bandwidth}}, but {{Not Both}}.
\newblock In {\em 2020 {{IEEE International Symposium}} on {{Performance
  Analysis}} of {{Systems}} and {{Software}} ({{ISPASS}})\/} (Boston, MA, USA,
  Aug. 2020), IEEE, pp.~180--191.

\bibitem{katsarakisHermesFastFaultTolerant2020}
{\sc Katsarakis, A., Gavrielatos, V., Katebzadeh, M.~S., Joshi, A., Dragojevic,
  A., Grot, B., and Nagarajan, V.}
\newblock Hermes: {{A Fast}}, {{Fault-Tolerant}} and {{Linearizable Replication
  Protocol}}.
\newblock In {\em Proceedings of the {{Twenty-Fifth International Conference}}
  on {{Architectural Support}} for {{Programming Languages}} and {{Operating
  Systems}}\/} (Lausanne Switzerland, Mar. 2020), ACM, pp.~201--217.

\bibitem{khalilov2024osmosis}
{\sc Khalilov, M., Chrapek, M., Shen, S., Vezzu, A., Benz, T., Di~Girolamo, S.,
  Schneider, T., De~Sensi, D., Benini, L., and Hoefler, T.}
\newblock $\{$OSMOSIS$\}$: Enabling $\{$Multi-Tenancy$\}$ in datacenter
  $\{$SmartNICs$\}$.
\newblock In {\em 2024 USENIX Annual Technical Conference (USENIX ATC 24)\/}
  (2024), pp.~247--263.

\bibitem{lamport1998parttime}
{\sc Lamport, L.}
\newblock The part-time parliament.
\newblock {\em ACM Transactions on Computer Systems (TOCS) 16}, 2 (May 1998),
  133--169.

\bibitem{lamport2001paxos}
{\sc Lamport, L., et~al.}
\newblock Paxos made simple.
\newblock {\em ACM SIGACT News 32}, 4 (2001), 18--25.

\bibitem{lamportVerticalPaxosPrimarybackup2009}
{\sc Lamport, L., Malkhi, D., and Zhou, L.}
\newblock Vertical paxos and primary-backup replication.
\newblock In {\em Proceedings of the 28th {{ACM}} Symposium on {{Principles}}
  of Distributed Computing\/} (Calgary AB Canada, Aug. 2009), ACM,
  pp.~312--313.

\bibitem{leisersonTheresPlentyRoom2020}
{\sc Leiserson, C.~E., Thompson, N.~C., Emer, J.~S., Kuszmaul, B.~C., Lampson,
  B.~W., Sanchez, D., and Schardl, T.~B.}
\newblock There's plenty of room at the {{Top}}: {{What}} will drive computer
  performance after {{Moore}}'s law?
\newblock {\em Science 368}, 6495 (June 2020), eaam9744.

\bibitem{li2016just}
{\sc Li, J., Michael, E., Sharma, N.~K., Szekeres, A., and Ports, D.~R.}
\newblock Just say $\{$NO$\}$ to paxos overhead: Replacing consensus with
  network ordering.
\newblock In {\em 12th USENIX Symposium on Operating Systems Design and
  Implementation (OSDI 16)\/} (2016), pp.~467--483.

\bibitem{li1989memory}
{\sc Li, K., and Hudak, P.}
\newblock Memory coherence in shared virtual memory systems.
\newblock {\em ACM Transactions on Computer Systems (TOCS) 7}, 4 (1989),
  321--359.

\bibitem{mahajan2011depot}
{\sc Mahajan, P., Setty, S., Lee, S., Clement, A., Alvisi, L., Dahlin, M., and
  Walfish, M.}
\newblock Depot: Cloud storage with minimal trust.
\newblock {\em ACM Transactions on Computer Systems (TOCS) 29}, 4 (2011),
  1--38.

\bibitem{marandi2010ringpaxos}
{\sc Marandi, P., Primi, M., Schiper, N., and Pedone, F.}
\newblock Ring paxos: A high-throughput atomic broadcast protocol.
\newblock In {\em Proceedings of the 2010 IEEE/IFIP International Conference on
  Dependable Systems and Networks (DSN)\/} (June 2010), pp.~527--536.

\bibitem{marvell2021liquidio}
{\sc {Marvell Technology Group Ltd.}}
\newblock {Multi-Core Processors - LiquidIO Smart NICs | Network Adapter}.
\newblock
  \url{https://www.marvell.com/products/infrastructure-processors/multi-core-processors/liquidio-smart-nics.html}.

\bibitem{michalowicz2023battle}
{\sc Michalowicz, B., Suresh, K.~K., Subramoni, H., Panda, D. K.~D., and Poole,
  S.}
\newblock Battle of the bluefields: An in-depth comparison of the bluefield-2
  and bluefield-3 smartnics.
\newblock In {\em 2023 IEEE Symposium on High-Performance Interconnects
  (HOTI)\/} (2023), IEEE, pp.~41--48.

\bibitem{netronome2021agilio}
{\sc {Netronome}}.
\newblock {Agilio CX 2x40GbE}.
\newblock
  \url{https://www.netronome.com/media/documents/PB_Agilio_CX_2x40GbE-7-20.pdf},
  2021.

\bibitem{nvidia_bluefield2}
{\sc {NVIDIA}}.
\newblock {NVIDIA BLUEFIELD-2 DPU}.
\newblock
  \url{https://www.nvidia.com/content/dam/en-zz/Solutions/Data-Center/documents/datasheet-nvidia-bluefield-2-dpu.pdf},
  2023.

\bibitem{ongaroSearchUnderstandableConsensus2014}
{\sc Ongaro, D., and Ousterhout, J.}
\newblock In search of an understandable consensus algorithm.
\newblock In {\em Proceedings of the 2014 {{USENIX}} Conference on {{USENIX
  Annual Technical Conference}}\/} (USA, June 2014), {{USENIX ATC}}'14, USENIX
  Association, pp.~305--320.

\bibitem{ousterhout15_ramcl_storag_system}
{\sc Ousterhout, J., Gopalan, A., Gupta, A., Kejriwal, A., Lee, C., Montazeri,
  B., Ongaro, D., Park, S.~J., Qin, H., Rosenblum, M., Rumble, S., Stutsman,
  R., and Yang, S.}
\newblock The {{RAMCloud}} storage system.
\newblock {\em ACM Transactions on Computer Systems 33}, 3 (2015).

\bibitem{palmieri2011osare}
{\sc Palmieri, R., Quaglia, F., and Romano, P.}
\newblock Osare: Opportunistic speculation in actively replicated transactional
  systems.
\newblock In {\em 2011 IEEE 30th International Symposium on Reliable
  Distributed Systems\/} (2011), IEEE, pp.~59--64.

\bibitem{petersen1996bayou}
{\sc Petersen, K., Spreitzer, M., Terry, D., and Theimer, M.}
\newblock Bayou: replicated database services for world-wide applications.
\newblock In {\em Proceedings of the 7th workshop on ACM SIGOPS European
  workshop: Systems support for worldwide applications\/} (1996), pp.~275--280.

\bibitem{pokeAllConcurLeaderlessConcurrent2017}
{\sc Poke, M., Hoefler, T., and Glass, C.~W.}
\newblock {{AllConcur}}: {{Leaderless Concurrent Atomic Broadcast}}.
\newblock In {\em Proceedings of the 26th {{International Symposium}} on
  {{High-Performance Parallel}} and {{Distributed Computing}}\/} (New York, NY,
  USA, June 2017), {{HPDC}} '17, Association for Computing Machinery,
  pp.~205--218.

\bibitem{RyabininGS24}
{\sc Ryabinin, F., Gotsman, A., and Sutra, P.}
\newblock Swiftpaxos: Fast geo-replicated state machines.
\newblock In {\em 21st {USENIX} Symposium on Networked Systems Design and
  Implementation, {NSDI} 2024, Santa Clara, CA, April 15-17, 2024\/} (2024),
  L.~Vanbever and I.~Zhang, Eds., {USENIX} Association, pp.~345--369.

\bibitem{redis}
{\sc Sanfilippo, S.}
\newblock Redis, 2023.
\newblock Accessed: 2023-01-22.

\bibitem{terraceObjectStorageCRAQ2009}
{\sc Terrace, J., and Freedman, M.~J.}
\newblock Object {{Storage}} on \{\vphantom\}{{CRAQ}}\vphantom\{\}:
  \{\vphantom\}{{High-Throughput}}\vphantom\{\} {{Chain Replication}} for
  \{\vphantom\}{{Read-Mostly}}\vphantom\{\} {{Workloads}}.
\newblock In {\em 2009 {{USENIX Annual Technical Conference}} ({{USENIX ATC}}
  09)\/} (2009).

\bibitem{renesseChainReplicationSupporting2004}
{\sc van Renesse, R., and Schneider, F.~B.}
\newblock Chain {{Replication}} for {{Supporting High Throughput}} and
  {{Availability}}.
\newblock In {\em 6th {{Symposium}} on {{Operating Systems Design}} \&
  {{Implementation}} ({{OSDI}} 04)\/} (2004).

\bibitem{vasilisgavrielatosantonioskatsarakisOdysseyImpactModern2019}
{\sc Vasilis~Gavrielatos, Antonios~Katsarakis, V.~N.}
\newblock Odyssey: {{The Impact}} of {{Modern Hardware}} on
  {{Strongly-Consistent Replication Protocols Vasilis}}.

\bibitem{vogels2009eventually}
{\sc Vogels, W.}
\newblock Eventually consistent.
\newblock {\em Communications of the ACM 52}, 1 (jan 2009), 40--44.

\bibitem{whittaker2021scaling}
{\sc Whittaker, M., Ailijiang, A., Charapko, A., Demirbas, M., Giridharan, N.,
  Hellerstein, J.~M., Howard, H., Stoica, I., and Szekeres, A.}
\newblock Scaling replicated state machines with compartmentalization.
\newblock {\em Proceedings of the VLDB Endowment 14}, 11 (2021), 2203--2215.

\bibitem{whittaker2020matchmaker}
{\sc Whittaker, M., Giridharan, N., Szekeres, A., Hellerstein, J.~M., Howard,
  H., Nawab, F., and Stoica, I.}
\newblock Matchmaker paxos: A reconfigurable consensus protocol [technical
  report].
\newblock {\em arXiv preprint arXiv:2007.09468\/} (2020).

\bibitem{xu14_charac_faceb_memcac_workl}
{\sc Xu, Y., Frachtenberg, E., Jiang, S., and Paleczny, M.}
\newblock Characterizing {{Facebook}}'s {{Memcached Workload}}.
\newblock {\em IEEE Internet Computing 18}, 2 (Mar. 2014), 41--49.

\bibitem{10.1145/3627703.3650071}
{\sc Zhou, Y., Wilkening, M., Mickens, J., and Yu, M.}
\newblock Smartnic security isolation in the cloud with s-nic.
\newblock In {\em Proceedings of the Nineteenth European Conference on Computer
  Systems\/} (New York, NY, USA, 2024), EuroSys '24, Association for Computing
  Machinery, p.~851–869.

\end{thebibliography}
